\documentclass{emulateapj}
\slugcomment{{\sc Accepted to AJ:} March 2, 2009} 
\usepackage{graphicx,natbib}

\newcommand{\FeII}{\ion{Fe}{2}}
\newcommand{\FeI}{\ion{Fe}{1}}
\newcommand{\HeI}{\ion{He}{1}}
\newcommand{\NIII}{\ion{N}{3}}
\newcommand{\LiI}{\ion{Li}{1}}
\newcommand{\CaI}{\ion{Ca}{1}}
\newcommand{\OI}{\ion{O}{1}}
\newcommand{\OIII}{\ion{O}{3}}
\newcommand{\SII}{\ion{S}{2}}
\newcommand{\HII}{\ion{H}{2}}

\shorttitle{Evidence for a Disk around MWC 778}
\shortauthors{Perrin, Vacca, and Graham}

\begin{document}

\singlespace

\title{Evidence for an Edge-On Disk around the Young Star MWC 778\\
from Infrared Imaging and Polarimetry}
\author{Marshall D. Perrin\altaffilmark{1}}
\affil{Division of Astronomy, University
of California, Los Angeles, CA 90095\\
mperrin@ucla.edu}
\author{William D. Vacca}
\affil{SOFIA-USRA, NASA Ames Research Center, Moffett Field, CA 94035\\
wvacca@sofia.usra.edu}
\and
\author{James R. Graham\altaffilmark{1}}
\affil{Dept.\ of Astronomy, University
of California, Berkeley, CA 94720-3411\\
jrg@berkeley.edu} 

\altaffiltext{1}{Center for Adaptive Optics, 
University of California, Santa Cruz, CA 95064, U.S.A. }

\begin{abstract}

MWC 778 is an unusual and little-studied young stellar object located in the IC 2144 nebula. Recent spectroscopy by Herbig and Vacca (2008) suggested the presence of an edge-on
circumstellar disk around it. 
We present near-infrared adaptive optics imaging polarimetry and mid-infrared imaging
which directly confirm the suspected nearly-edge-on disk around MWC 778 ($i \sim 70\degr - 80\degr$)
plus reveal a more extensive envelope pierced by bipolar outflow cavities.  
In addition, our
mid-infrared images and near-infrared polarization maps detect a spiral-shaped structure surrounding MWC 778, with 
arms that extend beyond 6\arcsec\ on either side of the star.

Although MWC 778 has previously been
classified as an Herbig Ae/Be star, the properties of its central source (including its spectral type) remain fairly
uncertain. 
Herbig \& Vacca (2008) suggested an F or G spectral type based on the presence of metallic absorption lines in the optical
spectrum, which implies that MWC 778 may belong to the fairly rare class of
Intermediate-Mass T Tauri Stars (IMTTSs) which are the evolutionary precursors to Herbig Ae/Be objects.
Yet its integrated bolometric luminosity, $\gtrsim 750 L_\odot$
(for an assumed distance of 1 kpc) is surprisingly high for an F or G spectral type, even for an IMTTS. 
We speculate on
several possible explanations for this discrepancy, including its true distance being much closer than 1 kpc, 
the presence of a binary companion, and/or
a non-stellar origin for the observed absorption lines.

\end{abstract}

\section{Introduction}

Most, if not all, low- and intermediate-mass stars form surrounded by
massive disks of gas and dust, which persist for a few million years before
dissipating. These disks play important roles in
regulating stellar angular momentum, launching bipolar winds and
outflows, and, for a substantial fraction of stars, giving rise
to planetary systems. 
Modeling a
disk's spectral energy distribution (SED) alone provides
only limited insight into these processes because young disks are optically thick, resulting in
degeneracies between model parameters and ambiguous interpretations
\citep{chiang2001}. 
Spatially resolved images in
scattered light are generally needed in order to break these degeneracies, and thereby probe the
disks' complex structures \citep{2007prpl.conf..523W}.

The two most well-known and well-studied classes 
of pre-main-sequence (PMS)
stars are the T Tauri and Herbig Ae/Be stars. 
The observational distinction between these two classes is based on spectral type: 
young stars cooler than F-type are considered T Tauris while hotter ones are
called Herbig Ae/Be stars.  
Typically, T Tauri stars are interpreted to be PMS stars of roughly solar mass or less, while
Herbig Ae/Be stars are young stars of intermediate mass, $2 M_\odot < M < 8
M_\odot$. 
But such a one-dimensional classification 
cannot fully represent the range of PMS stellar
properties, which depend on both mass and age. 
This is
particularly true for intermediate-mass stars, which are stable against convection
and consequently evolve horizontally across the
Hertzsprung-Russell diagram towards the main sequence, rising in
surface temperature and changing in spectral type as they go. 
As very young intermediate-mass stars contract toward becoming Herbig Ae/Be stars,
they first pass through a stage in which they are classified as T Tauri stars, but ones with
vastly higher luminosity than typical. These ``intermediate-mass T
Tauri stars'' (IMTTS; Calvet et~al. 2004) are the youngest
intermediate-mass stars available for observation at visible or near-IR
wavelengths. Due to the slope of the stellar initial mass function and
the rapid speed at which stars evolve through this early stage, IMTTSs are
relatively rare and have been the subject of few detailed studies. Yet
these objects potentially offer insights into many aspects of the star
formation process, such as the dissipation of stars' natal
molecular clouds and the properties of circumstellar disks at very
young ages. 

MWC 778 ( = IRAS 05471+2351) is an unusual, and little-studied, object 
located in the IC 2144 nebula in the direction of the Galactic anticenter. 
Herbig \& Vacca~(2008;~hereafter HV08) recently presented high
resolution optical and moderate resolution 
near-infrared spectra of MWC 778 and its surrounding nebulosity.
These spectra suggest that MWC 778 is a mid- to late-type PMS star with a
bolometric luminosity of several hundred $L_\odot$ --- in other words,
it appears that it may be an IMTTS, though this classification is far
from certain, as we will discuss below.
Their spectra furthermore imply the presence of a circumstellar disk around 
MWC 778, inclined close to edge-on to our line of sight. 

In this paper, we present near-IR adaptive optics (AO) imaging
polarimetry and mid-IR imaging that resolve this disk,
confirming the interpretation of HV08.  These observations were taken
as part of a larger survey of Herbig Ae/Be stars with adaptive optics
polarimetry and mid-infrared imaging (Perrin 2006).
Near-IR differential polarimetry is an effective technique
for resolved imaging of circumstellar material because it allows
suppression of the extended and time-variable AO point spread function
(PSF) halo.  Because atmospheric turbulence does not polarize the PSF
halo, while dust-scattered light is strongly polarized, differential
polarimetry can separate them to reach the fundamental photon noise
limit for detection of faint material
\citep[e.g.][]{2001ApJ...553L.189K,2004AnA...415..671A,
Perrin:2008p2351}.

The following section \S \ref{previous} summarizes earlier studies of
MWC 778 and briefly recaps the results of HV08. In \S
\ref{obs_sect}, we describe our observations: near-IR imaging
polarimetry from Lick Observatory and mid-infrared imaging from Gemini
North, and \S \ref{results_sect} presents the results of these observations.
In \S 5 we discuss the implications of these data for the nature of
MWC 778 itself and its surrounding nebula, and in particular consider
several competing hypotheses for the spectral type and luminosity of the central
illuminating source.  \S \ref{concl_sect} summarizes our
conclusions.

\section{ Previous Studies of MWC 778 }
\label{previous}

First
identified in the Mount Wilson Catalogue of H$\alpha$-emitting objects
compiled by Merrill \& Burwell (1949), MWC 778 
was classified as Bpe based on a low-dispersion slit spectrogram
(obtained by Minkowski) which revealed strong H and \FeII\ emission
lines. Allen (1974) noted that MWC 778 has a large infrared excess due to
dust emission, giving it a very unusual SED that rises nearly linearly in a
plot of $\lambda F_\lambda$ vs. $\lambda$ between 1.65 and 18 microns.
This rising infrared SED places MWC 778 firmly within the category of
Class I YSOs, with a NIR spectral index $\alpha \sim 0.4$. 
Allen also reported the presence of emission lines of H, \FeII, [\FeII],
and [\SII] in an optical spectrum of the source, and suggested that it
was a pre-main sequence Ae or Be star.
	
	Garc\'{i}a-Lario et~al.\ (1997) classified MWC 778 (=  PDS 204 = GLMP~131) as a YSO based on its 
	near-infrared and IRAS colors. They suggested it was associated with the Galactic \HII\ region BFS50 (= IC 2144), and
	classified it as a BQ[ ] (which is equivalent to B[e] in a more modern classification system) based on
	the similarity of its moderate-resolution optical spectrum to that
	of HD 51585 and other B[e] stars. This spectrum (shown as Fig.\ F.\ 1 in  Su\'{a}rez et~al.\ 2006) exhibits
	numerous permitted and forbidden emission lines, most of which are attributed to \FeII. 
		
	Vieira et~al.\ (2003) obtained $UBVRI$ photometry of MWC 778 as well as moderate resolution (R=9000) optical
	spectra centered near H$\alpha$. Although the optical SED was found to be quite red in general (as might be
	expected for a source with such a large infrared excess), MWC 778 exhibits a very blue $U-B$ color  ($U-B= -0.27$).
	The H$\alpha$ line was seen to be double-peaked with unequal strengths in the two peaks. They classified the
	object as B1? and noted the presence of [\OI] and [\SII] in emission as well. They associated MWC 778 with the
	$^{12}$CO ($J=1-0$) emission source in the star-forming region WB711 (Wouterloot \& Brand 1989), for which
	Wouterloot et~al.\ (1993) gave a distance of 1.1 kpc and a far-infrared luminosity $L_{FIR} = 400 L_\odot$
	(derived from the IRAS fluxes).  
				They noted that the integrated luminosity resulting from the
	SED combined with the assumed distance led to a location in the HR diagram for MWC 778 that was ``out of the
	expected position for HAeBe stars".
	
	Herbig \& Vacca (2008; hereafter HV08) recently presented high resolution optical (R=48000) and moderate
resolution near-infrared (R=2000) spectra of MWC 778, as well as a deep image taken through an H$\alpha$ filter.  They inferred the presence of
a circumstellar disk based on the double peaked nature of many of the emission lines. These authors examined the optical
spectrum of Vieira et~al.\ (2003) and questioned the reported B spectral type classification. Their own much-higher-resolution optical spectra reveal the presence of rotationally broadened absorption lines from species like
Ca I and Fe I, which they claimed were photospheric in origin and based on which they suggested a spectral type for MWC 778 between late A and G. The NIR spectrum exhibits a
large number of \FeII\ and [\FeII] lines as well as the Br series (up to Br 26) and the Pa series (up to Pa 23) in
emission. Herbig \& Vacca derived an extinction $A_V \sim 2-3$ from a variety of indicators and estimated a distance of $\sim 1$
kpc based on associations with other star forming regions in the area, but stressed that this value is highly
uncertain. Integrating the observed SED using that distance leads to a bolometric luminosity of  $\sim 510 L_\odot$. The
H$\alpha$ image shown by HV08 looks very similar to the lobed nebula surrounding other B[e] stars (Marston \& McCollum
2008). HV08 refer to IC 2144 as a reflection nebula, not an emission nebula (i.e., \HII\ region), though it has also been
classified as an \HII\ region by others (e.g. \citealp{blitz82}).

\section {Observations and Data Reduction}
\label{obs_sect}

\begin{figure*}[t]
\includegraphics[height=3.70in,keepaspectratio=true,origin=c]{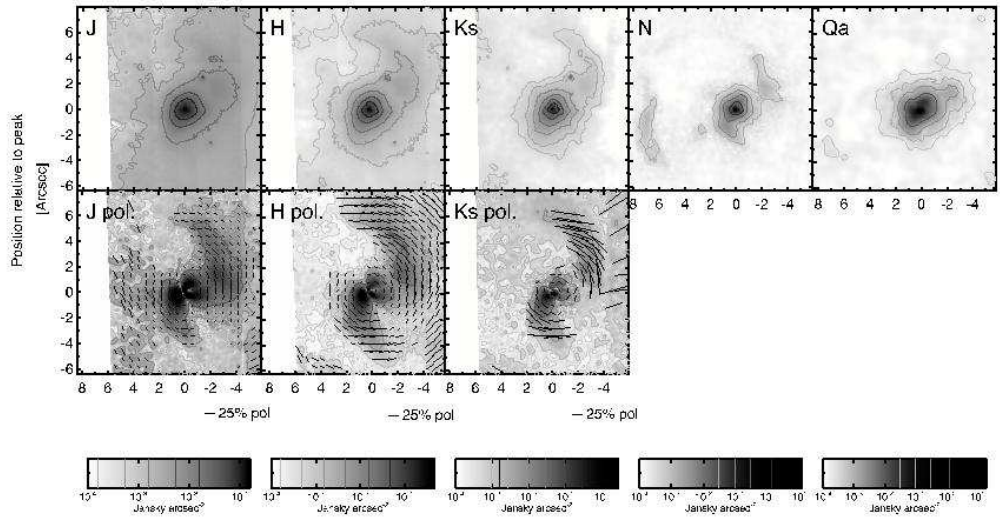}
\caption{\textbf{Top row:}  
Total intensity near- and mid-infrared images of MWC 778. The display scales for each wavelength are shown below, with contour levels as indicated.  \textbf{Bottom row:} 
Polarized intensity images in $J$, $H$, and $K$ bands, with vectors showing polarization fraction and orientation. The display scales for polarized intensity are reduced 100x relative to the scales for total intensity shown in the scale bars.
North is at the top, east to the left in all images. 
Immediately around MWC 778, the near-IR polarization maps reveal two bright lobes bisected by a narrow dark lane with a position angle of $\sim
25 \degr$ (See also Figure 2 for a closer view). 
This region is embedded in a larger spiral-shaped nebula, which can be seen extending to at least 6\arcsec\ radius on
either side of the star.  It is particularly visible in the $N'$ image, where the northwest spiral arm can be seen to
wrap 180$\degr$ around the star. \\
}
\label{Fig. 1}
\end{figure*}

\subsection{Near-Infrared Imaging Polarimetry}

We obtained $J, H$ and $K_s$ band differential imaging linear polarimetry of MWC 778 with the 3-m Shane telescope at Lick
Observatory, using the laser guide star adaptive optics (LGS AO) facility  \citep{Gavel:2003p2513} and the IRCAL camera
\citep{Lloyd00}.  

A detailed description of the IRCAL polarimeter, the observing procedures, and the associated data
reduction pipeline can be found in Perrin et~al.\ (2008). 
Briefly, a Wollaston prism within IRCAL forms simultaneous images of perpendicular polarizations on the detector, and a rotating
half-wave plate modulates the observed polarization state while a sequence of images are taken.  After bias and flat-field correction,
the images for each linear polarization (corresponding to a given waveplate rotation) are mosaiced together. The sums
and differences of the resulting mosaic images then give the Stokes $I, Q$ and $U$ parameters describing linear
polarization.  
To improve sensitivity to faint signals, the reduced images presented in this paper were adaptively smoothed by convolution with a
3-resolution-element wide Gaussian in regions with signal to noise (S/N) $<$ 1 per pixel, decreasing to one resolution element wide for $1
< $S/N $< 10$, and no smoothing for pixels with S/N $>$ 10.  The near-IR total intensity images are shown in Figure 1 (top
row), along with the linear polarized intensities $P = (Q^2 + U^2)^{1/2}$, where $Q$ and $U$ are the usual Stokes
parameters. 

These observations were taken on 2005 Nov 23, during a period of unseasonably warm weather and good seeing: Fried's
parameter $r_0$ was 15-20 cm that night.  For MWC 778, our total exposure times in $J$, $H$, and $K_s$ bands were 720 s, 640 s, and 390 s,
respectively.  The achieved point spread function (PSF) full width at half maximum (FWHM) was 0.29\arcsec\ at $J$, 0.21\arcsec\ at $H$, and
0.20\arcsec\ at $K_s$, as measured from the star 2.8\arcsec\ northwest of MWC 778. Because MWC 778 itself is resolved, we cannot measure PSF Strehl ratios directly on it, and the nearby star to the northwest is also unsuitable due to the spatially-variable bright nebulosity behind it. As a proxy, we measured\footnote{Using the \texttt{lickstrehl} IDL tool; 
http://mthamilton.ucolick .org/techdocs/instruments/ircal/ircal\_lickstrehl.html.
See \citet{2004SPIE.5490..504R} for a discussion of the difficulty in measuring Strehl ratios accurately for undersampled instruments such as IRCAL. The algorithm in \texttt{lickstrehl} incorporates the best practices identified by Roberts et al., but we caution there may still be systematics particularly at these low correction levels.}
the Strehl ratio for an unresolved star of comparable brightness observed with LGS AO immediately after MWC 778 (PDS 211, $V$=13.7), and found Strehl ratios of $0.02\pm0.01$, $0.07\pm0.02$ and $0.16\pm0.03$ for $J, H$ and $K_s$ respectively, corresponding to 400-500 nm rms wavefront error.
For photometric calibration we 
observed the standard S852-C \citep{Persson:1998p951} shortly after MWC 778.  

Our images reveal a bright point source surrounded by 
    fainter extended nebulosity filling most of our 12\arcsec$\times$20\arcsec\ field of view
    (see \S \ref{results_sect}). 
We performed
aperture photometry around MWC 778 using both 1\arcsec\ and 6\arcsec\ radius apertures to estimate the flux from the central
bright region alone and the whole observed nebula, respectively (Table \ref{photom_table}).  There were intermittent
cirrus clouds that night, but we observed MWC 778 during a clear period, and our derived photometry is in excellent
agreement with that of HV08 taken one month earlier, suggesting that the cirrus did not adversely impact these data.  We caution that even
the 6\arcsec\ radius does not include the \textit{entire} nebula,
only the portion visible within our 12\arcsec-wide field of view. In
the 2MASS Atlas images, the nebula around MWC 778 is seen to
extend at least 20\arcsec\ from the star, and in deeper images would
likely appear comparable in size to the
$H\alpha$ nebula shown in HV08, $\sim 2$ arcmin across.

\begin{figure*}
\includegraphics[height=4.5in,keepaspectratio=true,origin=c]{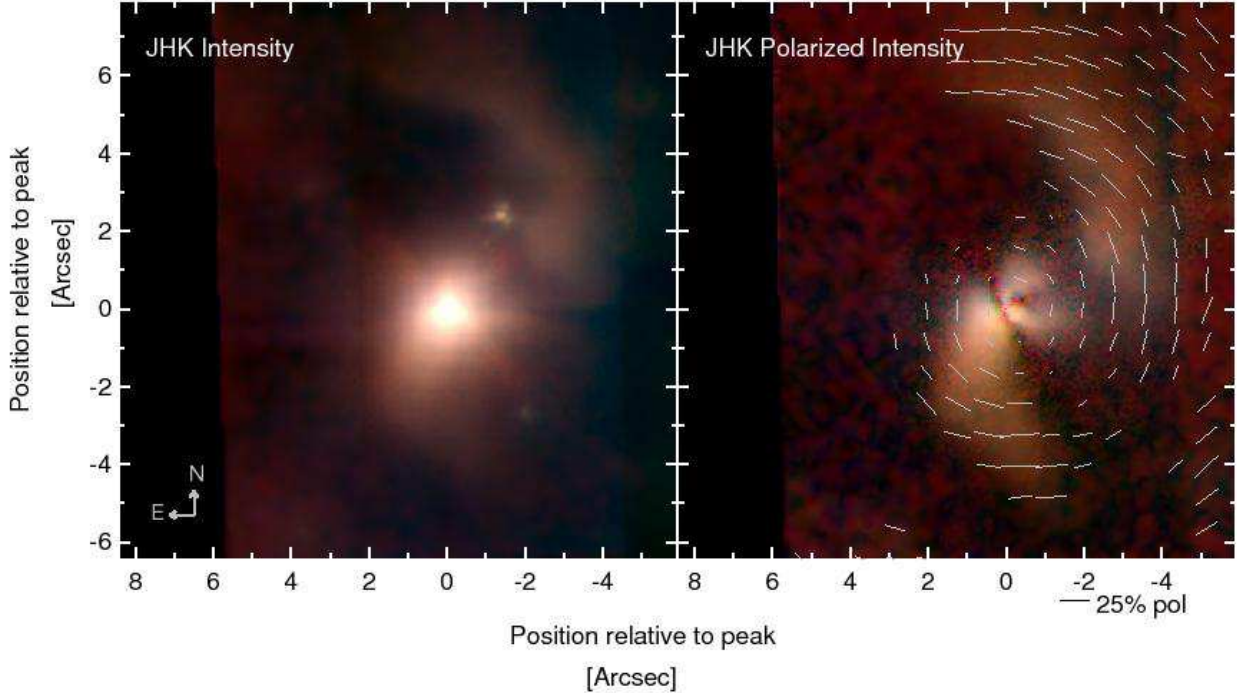}
\caption{ IRCAL 3-color near-infrared image of the total intensity and polarization in MWC 778. The overplotted
vectors show the polarization position angle and fraction as measured in $H$ band.
The observed dark lane is the characteristic signature of a nearly edge-on circumstellar disk,
while the partially-divided northwest nebulosity suggests the limb-brightened edges of a conical cavity in a circumstellar
envelope, cleared by bipolar outflows.  
The location of peak brightness in total intensity (at left) 
is displaced
from the center of the polarization vectors by 0.15\arcsec, and at right corresponds in location 
to the bright region immediately east of the dark lane's middle. This displacement indicates that we do not see the star
directly at these wavelengths, but rather observe it entirely in light scattered around the circumstellar disk (See \S
\ref{results_disk})
}
\label{Fig. 2}
\end{figure*}

\subsection{Mid-infrared Imaging}

Mid-infrared images of MWC 778 at $11.2$ and $18.1$ microns ($N'$ and $Qa$ bands) were obtained at
Gemini North on 2005 Dec 16 using the Mid-Infrared Imager/Echelle Spectrometer (Michelle) instrument 
\citep{Glasse:1997p2517} as part of program GN-2005B-C-4. The total on-source integration times were 235 s and 218 s, respectively. The chop throw was $15"$ at a position
angle of $30\degr$. Conditions were clear, with good seeing and 2.4 mm of precipitable water vapor.

These data were reduced using custom IDL code that incorporates and extends IDL reduction routines provided by
R.~S.~Fisher at Gemini. A detailed description of this reduction code can be found in Perrin (2006). To remove the thermal background emission, each chop-nod set was reduced using the standard
double subtraction method. No flat fielding was done. All chop-nod-subtracted images were registered via Fourier
cross-correlation and co-added to generate a final mosaic. The data were then corrected for atmospheric transmission and
flux-calibrated via observations of standard stars taken from \citep{Cohen:1999p243} obtained throughout the night. 
 
The final images have PSF FWHMs of 0.39\arcsec\ at $N'$ and 0.54\arcsec\ at $Qa$, as measured on a calibration star
observed
immediately after MWC 778. The field of view is $32\arcsec\times 24\arcsec$. To improve sensitivity to faint signals, these images were adaptively smoothed by
convolution with a 3-resolution-element wide Gaussian in regions with S/N $<$ 1 per pixel, decreasing to one resolution
element wide for $1 <$ S/N $< 10$, and no smoothing for pixels with S/N $>$ 10. Photometry was performed on the
unsmoothed data, using for consistency the same 1\arcsec\ and 6\arcsec\ apertures as were used for the near-IR data. 
(See Table \ref{photom_table}).
We note that the emission observed in $N'$
actually extends beyond the 6\arcsec\ aperture, primarily in the form of an
arc of emission east of MWC 778 just outside the aperture. There is approximately 0.3 Jy
of emission from that region in $N'$. Along the chopping position angle of 30$
\degr$, no mid-IR nebulosity is visible beyond $\sim 7\arcsec$ radius at our sensitivity,
so the 15\arcsec\ chop throw does not result in any self-subtraction of the nebula.

\begin{figure*}
\includegraphics[height=3.5in,keepaspectratio=true,origin=c]{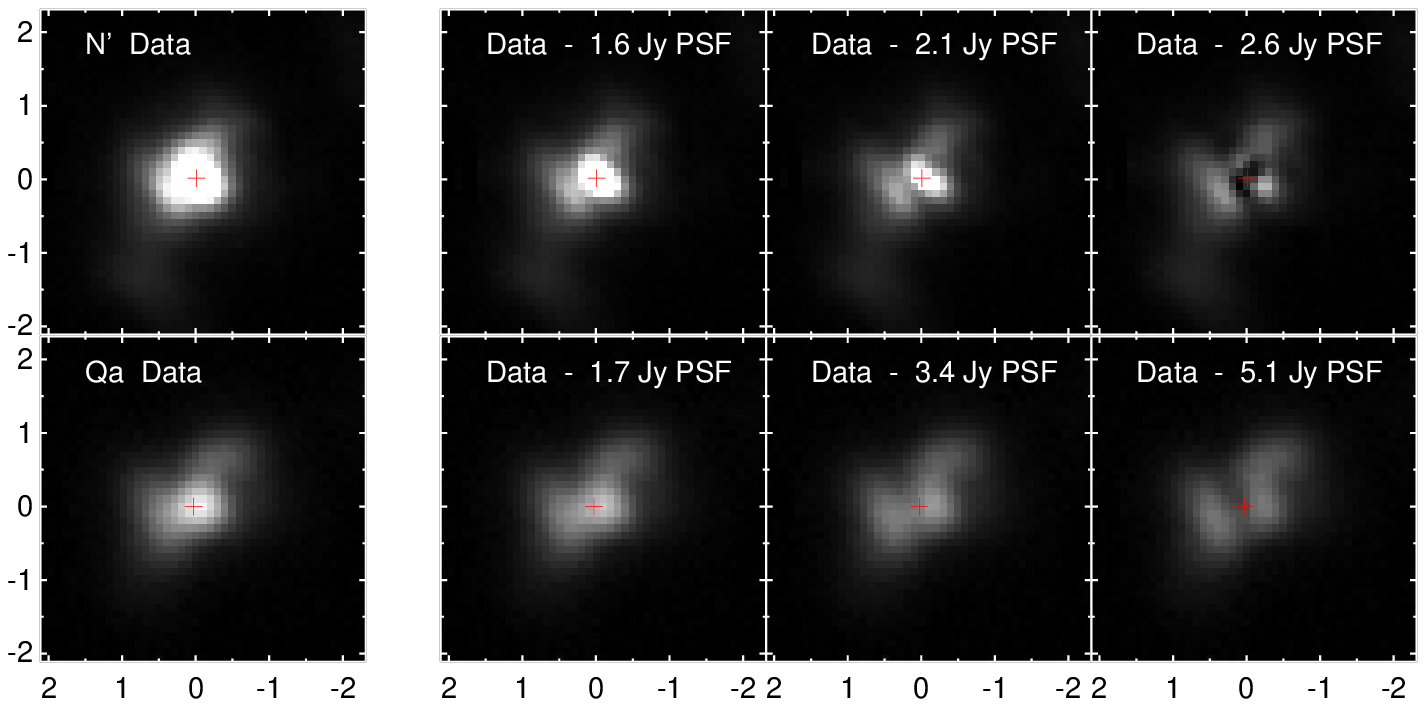}
\caption{\label{mirsub}Mid-infrared PSF subtractions, showing a mid-IR counterpart to the dark lane seen in the near-IR.
The unsubtracted observations are shown at left, while the right panels show PSF subtractions with various flux scalings
for the unresolved component. The cross symbols mark the central peak
location of the subtracted PSFs. 
For each wavelength, all four images are shown with the same linear display scaling.
The optimal PSF fluxes, 2.1 Jy for $N'$ and 3.4 Jy for $Qa$, were determined empirically based on interactive adjustment of the subtractions to
eliminate the central point source and produce smooth residuals; there is considerably uncertainty in their values. Yet the detection of the dark lane is a robust feature of
these data, persisting even if the subtracted PSF varies in intensity by $\pm25$\% at $N'$ or 50\% at $Qa$, as shown in the
adjacent panels.  The mid-IR dark lane 
appears at the same position angle and with comparable spatial extent to the near-IR dark lane shown in Figure 2, and the midpoint of the mid-IR dark lane is likewise offset $\sim 0.2\arcsec$ southeast of the location of peak flux. The asymmetric nebula
on either side of the dark lane also shows a similar bipolar structure consistent with the near-IR. \\
}
\end{figure*}

\section {The Complex Environment of MWC 778}
\label{results_sect}

\subsection{A spatially-resolved, nearly-edge-on disk and bipolar
envelope}

\label{results_disk}

Immediately around MWC 778, the near-IR polarization maps reveal two bright lobes bisected by a narrow dark lane with a position angle of $\sim
25 \degr$.  The dark lane is slightly curved, with the concave side facing northwest. On that side, the bright northwest nebulosity further splits into two fans of light separated by a region of
lower polarization.  The polarization vectors are centrosymmetric throughout, indicating that the observed structure is a reflection nebula
illuminated by MWC 778. This region is embedded in a larger spiral-shaped nebula, which we discuss below;
here we concentrate on the central portion.

The observed dark lane is the characteristic signature of a nearly edge-on circumstellar disk,
while the partially-divided northwest nebulosity suggests the limb-brightened edges of a conical cavity in a circumstellar
envelope, cleared by bipolar outflows.  Overall, MWC 778 bears a striking resemblance to several other young stars known
to have nearly-edge-on disks and envelopes with bipolar cavities, for instance LkH$\alpha$ 233 (Perrin et~al. 2004a).
Based in part on discrepancies between optical spectra of MWC 778 and the nearby nebulosity, HV08 inferred the
presence of an inclined disk around the star, with its rotational axis pointing to the northwest. We confirm here that
suggestion via direct detection of the disk in our polarization maps. The observed rotational
axis of the disk agrees with that suggested by HV08.

The position of peak intensity in the total intensity images is
shifted about 0.2\arcsec\ (1 resolution element) northwest from the center of the dark lane. 
This displacement toward the northwest
indicates that side of the disk is tipped slightly toward us. The slight curvature of the central dark lane confirms
this interpretation: MWC 778's disk is nearly edge-on, but not perfectly so,
or it would obscure the central star entirely and the dark lane would be straight. The precise amount of inclination consistent with the observations depends on the disk's structure, such
as the amount of flaring, which we currently cannot constrain. However, it seems probable that the disk is inclined
roughly 70-80\degr\ to our line of sight, such that the star is just
barely visible around the obscuring disk. The central intensity peak is 
significantly polarized, $1.8\% \pm 0.4\%$ at $H$, which suggests that
in fact we see the star in light scattered around the disk's edge rather
than directly.

The apparent polarization fraction rises both at longer wavelengths and with increasing distance from the star. For instance, in the bright part of the southeastern nebula,
about
1.5\arcsec\ from the star, the $J, H, Ks$ polarization fractions are $\sim 13\%, 16\%,$ and $18\%$ respectively, while in the curved
region 4\arcsec\ northwest of the star, the polarization fractions are $22\%, 28\%$ and $34\%$. However, within the
central arcsecond on both sides of the dark lane, the polarization fraction stays approximately constant with wavelength at
$\sim9\%$. At face value, this may suggest different dust grain properties closer to the star, as would be expected from collisional aggregation of larger particles. But 
it is hard to accurately measure polarization fractions with AO data, given the contamination
from uncorrected stellar light in the AO PSF halo, so these numbers should be considered cautiously.

\begin{deluxetable*}{lr|rrrr|rlrl|}
\tablecolumns{6}
\tablewidth{0pt}
\tablecaption{\label{photom_table}Photometry of MWC 778}
\tablehead{
Band     & Wavelength & $R \leq 1\arcsec$ & Uncert. 	&  $R \leq 1\arcsec$ & Uncert. & $R \leq 6\arcsec$ & Uncert.     &  $R \leq 6\arcsec$ & Uncert. \\
         &            & (mag.)           &  (mag.)    	&  (Jy.)    &   (Jy)       & (mag.)             &              &  (Jy)              &        
}
\startdata   $J$ 	&  1.20		& 10.37	& 0.06	& 0.111	&0.006	& 10.15	& 0.06	& 0.136	& 0.007	\\	
$H$ 	&  1.60		&  9.31	& 0.05	& 0.197	&0.010  &  9.12	& 0.05	& 0.235	& 0.012	\\	
$K_s$ 	&  2.05		&  8.25	& 0.05	& 0.323	&0.016  &  8.04	& 0.05	& 0.392	& 0.020	\\	
$N'$ 	&  11.20	&  2.27	& 0.03	& 3.68~~&0.076  & 1.93	& 0.03 	& 5.08	& 0.12	\\	
$Qa$ 	&  18.10	& -0.50 & 0.03	&18.56~~&0.55	&-0.91	& 0.03	& 27.20 & 0.60	\\	
\enddata
\tablecomments{ Photometry of MWC 778, in both 1\arcsec\ and 6\arcsec\
apertures. The measurements are presented as both magnitudes
and fluxes. The stated uncertainties include both
statistical error and systematic flux calibration uncertainty,
conservatively estimated at 5\% in the near-IR and 1.5\% in the mid-IR. (The more
stable mid-IR PSF allows more precise absolute calibration compared to the
near-IR AO PSF.) Note that the smaller 1\arcsec\ aperture still
contains some amount of resolved nebular flux in addition to the central
point source.
}
\end{deluxetable*}

Both our mid-IR images show a bright point source at the location of
MWC 778, surrounded by extended nebulosity aligned approximately
southeast to northwest. In the $Qa$ image, a hint of the dark lane can
be seen, but it is partially obscured by the bright central point
source. We therefore subtracted the central source using scaled copies
of a PSF reference star observed immediately after MWC 778 (Figure
\ref{mirsub}). The PSF was aligned with the position of the central
peak as measured by fitting 2D Gaussians to both, and the overall flux
scaling was iteratively adjusted to remove the central peak without
oversubtracting the surrounding nebulosity. The optimal flux scalings
found this way are not particularly precise (and in fact are somewhat
subjective), but the basic result holds true for a wide range of PSF
scalings, as shown in Figure \ref{mirsub}: Subtracting the central PSF
reveals a dark lane in both the $N'$ and $Qa$ images, with the same
orientation and scale as that seen in the near-IR.  The optical depth
in the disk midplane must be sufficient to cause appreciable
extinction even at 18 $\mu$m, implying $A_V > 50$ mag.\ in visible
light in the disk midplane. The light we see at short wavelengths must
scatter out around the disk through paths of much lower extinction. 

We caution that the point-source fluxes found in this manner, 2.1 Jy for $N'$ and
3.4 Jy for $Qa$, almost certainly do not represent the photospheric
intensity of the central star. Thermal emission from a circumstellar
disk on scales of $\lesssim 100$ AU would be unresolved by these
observations. The observed flux levels are far in excess of the expected photospheric flux of any plausible central star,
leading us to conclude that in these point sources we are most likely observing thermal emission from the hot inner disk.

The dark lane extends outward for $\sim 1.2 \arcsec$ on either side of the star in the near-IR, and almost as far in the mid-IR, corresponding to a
physical radius of $\sim 1200$ AU for an assumed distance of 1 kpc. This is similar in size to other
spatially-resolved disks around Herbig Ae/Be stars (e.g. 800 AU for HD 141569, \citealp{Clampin:2003p118}; 1000 AU for LkH$\alpha$ 233, \citealp{Perrin:2004p1832}; 1100 AU for MWC 480, \citealp{Simon:2000p2528}).
The thickness of the dark
lane increases with radius from the star: in the near-IR close to the center, the apparent thickness is $\sim 0.2\arcsec$ (about 1
resolution element), but at a radius of 0.7\arcsec\ the thickness has increased to 0.5-0.6\arcsec.  

Does the observed extent of the dark lane
correspond to the physical diameter of the disk, or are we 
instead seeing a large shadow cast by a much smaller disk?  In their detailed
study of disk shadows, Pontoppidan et al. (2005) wrote that the ``most
important difference between dark lanes due to extinction from a large
disk and the dark lanes due to projection is that the former are
expected to be entirely dark while the latter will still have a bright
source in the center of the system... [For a disk shadow falling on a screen] a dark
band should be seen but
with a central near-infrared compact source in the center. The presented 
model never produces a disk shadow without the central source being at 
least as bright as the reflecting material.'' In the case of MWC 778,
no bright point source is visible \textit{within} the dark lane at any
wavelength. Instead the bright PSF peak is
offset northwest outside of the dark lane at all wavelengths, consistent with light scattering
around the edge of an optically thick disk. Hence we conclude that the
observed morphology is primarily due to a physically large disk with
apparent angular
radius $\sim 1''$,
rather than shadowing from a much smaller disk.

\subsection{A giant spiral?}

\label{spiral} On larger scales the polarized intensity maps and
mid-infrared images exhibit a pair of curving ``spiral arms'' that
wrap counterclockwise around the central disk.  This is an infrared
counterpart to the ``reverse question mark''-shaped dark lane HV08
saw\footnote{We note that another spiral structure can also be seen on
a vastly larger scale in HV08's H$\alpha$ image (see their Fig.\ 1).
This image reveals a faint, diffuse, ``spiral'' feature beginning
almost due west (PA=270\degr) of MWC 778 and extending nearly an
arcminute from MWC 778 while wrapping 90\degr\ around the source to
end at PA=0\degr. It is not clear if there is any connection between
the two spiral features.} in their H$\alpha$ image; the spiral arms we
see in the near- and mid-infrared are coincident with the bright
outline of the ``question mark'', while the darker region inwards of
the spiral forms the ``question mark'' itself.  Although this overall
structure can be discerned in total intensity images, particularly in
our $K_s$ and $N'$ images and the H$\alpha$ image of HV08, it is
especially prominent in our linear polarization maps (Figure 1). 

The spiral is asymmetric: the southeastern arm is both shorter and more tightly curved than
the northwestern arm. It appears to have a distinct corner about 3.5\arcsec\ southwest of the star, a sharp
feature also clearly seen in the $N'$ image. 
The southeast arm trails off about 6\arcsec\ from the star at PA=195\degr, while
the opposite northwest arm fades out in our near-IR images around $r=7\arcsec$, PA=15\degr. 
That places its apparent termination almost exactly 180\degr\ opposite the end of the southern arm, though a bit more distant from the star. 

Our mid-infrared images and the
H$\alpha$ images of HV08 (see their Fig.\ 3) show that this ``northwestern'' spiral arm in fact continues onward to wrap nearly $180 \degr$ around the source, 
albeit at reduced brightness for much of that arc.
It becomes brighter again in a region $\sim 6''$ to the E of MWC 778, seen 
prominently in the H$\alpha$, $N'$, and $Qa$ images. This bright region is mostly
outside the field of view of our near-IR images, but can partially be
seen at the extreme southeast corner of the polarized intensity maps. Curiously, near the base of the northwestern arm a
partial gap can be seen, a region of reduced polarized intensity between $1.5-2.4$\arcsec\ from the star. No similar gap
is seen in the southeastern arm.  

We discuss possible explanations for this spiral structure in \S 5.2 below.

\subsection{Detection of Candidate Companion Stars}

HV08 noted that if MWC 778 is indeed a massive YSO, then we should expect it to be surrounded by a cluster of lower-mass
YSOs. In our AO data, 
two fainter stars are visible near MWC 778; see Figure 2, where these point sources are apparent at roughly
(-2\arcsec, -2\arcsec) and (-1.5\arcsec, 2\arcsec) from MWC 778. 
Table \ref{comps} presents the photometry and astrometry of
these potential companions. 
Neither of these stars are visible in our mid-IR data. 
It remains unknown whether these are physically associated young stars
or merely foreground objects. Given the amount of circumstellar dust around MWC 778, it seems unlikely for them to be background objects. 

In addition, in our $N'$ image, two other faint point sources ($\Delta{N'} \sim 7$ mag.)
are detected several arcseconds southwest of MWC 778. Both of these
objects fall outside the field of view of our near-IR data (and consequently are not shown in Figure 1, where we have cropped the
mid-IR images to the same FOV as the near-IR data). However,
their positions match known point sources from the
2MASS catalog. These two sources are also included in Table \ref{comps}. 

On the other hand, the 2MASS ``point sources'' 2MASS 05501321+2352179, 05501322+2352242, 05501418+2352115, 05501438+2352132, and
05501443+2352175 correspond to the resolved nebulosity around MWC 778. We
see no evidence in our data of actual point sources at those
locations.

\begin{deluxetable*}{lrrrrrr}
\tablecolumns{6}
\tablewidth{0pt}
\tablecaption{\label{comps}Nearby Sources}
\tablehead{
 Source        &  $\rho$ & $\theta$ & $\Delta J$  &$\Delta H$ & $\Delta K_s$ & $\Delta N'$ 
}
\startdata
      MWC 778 NIR1 &   2.81 &  330.6 &   4.22 &   4.27 &   4.83  & - \\ 
      MWC 778 NIR2 &   3.29 &  216.9 &   5.65 &   6.24 &   6.99  & -\\ 
2MASS 05501333+2352136 &   8.95 &  239.4 &     -  &     -  &     -   &  7.32 \\ 
2MASS 05501303+2352140 &  13.08 &  253.5 &     -  &     -  &     -   &  7.09\\ 
\enddata
\tablecomments{Properties of other sources detected in our fields of view.  The
two near-IR sources were not detected in the mid-IR data, while the
$N'$ point sources are outside the field of view of our near-IR data.
Astrometry is relative to MWC 778, and photometry at each wavelength is relative to the observed flux of MWC 778 measured within
a 1\arcsec\ aperture, though the photometry of these stars used
0.5\arcsec\ apertures to reduce contamination from the nebula.
Aperture corrections were applied based on subsequent observations of
photometric standard stars. Derived uncertainties are $\Delta{\rho}
=\pm 0.05\arcsec$, $\Delta\theta=1\degr$, and $\sim 5\%$ for the
relative photometry. The near-IR point sources are both visible near MWC 778
in Figure 2. The mid IR point sources are not shown in any of the figures here due to the field of view selected for
display.}
\end{deluxetable*}

\section{Discussion}
\label{discussion_sect}

HV08 presented optical spectra that showed MWC 778 and its nearby nebulosity
differ strikingly in which emission lines are present and in the 
shapes and velocities of those lines. 
They suggested that these spectral discrepancies could be explained if the star was seen through a rotating edge-on disk, while the portion of the nebula they
observed (northwest of MWC 778) was directly illuminated by the star through a polar cavity perpendicular to the disk
plane. Our observations resolve the disk and bipolar envelope with precisely the orientation claimed by HV08, and confirm their interpretation is correct.

But many other questions still remain open about MWC 778. 
One of the most fundamental questions is the distance, which remains very uncertain.  
The value of 1 kpc adopted by HV08 was based on 
distances to a handful of other clouds and star forming regions in that part of the sky, ranging from 0.4 to 2 kpc,
despite the fact that MWC 778 is not known to be physically associated with any of
those clouds.
This uncertainty in the distance to MWC 778 
impacts our ability to assess its physical properties and evolutionary state.  

\subsection{IC 2144 is a Reflection Nebula}

IC 2144, the nebula around MWC
778, has previously been described as both a reflection and an emission nebula (see \S 2). HV08
considered IC 2144 to be a reflection nebula, but did not explicitly state their
reasoning behind this classification. Three factors contributed to
their classification (G. Herbig, private communication 2008): 1. None of their high-resolution optical
spectra show any signs of the [\OIII] $\lambda\lambda
4959, 5007$ emission lines, which are ubiquitously found in \HII\
regions. 2. The H$\alpha$ emission line seen from the nebula 
has FWHM $\sim 3$ \AA, as broad 
as the H$\alpha$ emission from MWC 778 itself, and in 
contrast to the generally narrow emission typically seen from \HII\ regions (FWHM $\sim 0.5$ \AA). 
3. The spectrum of MWC 778 itself shows absorption lines indicative
of a relatively cool spectral type (e.g. \LiI, \CaI, \FeI), and lacks 
any absorption lines (e.g. \HeI\ and \NIII) that would indicate the presence of an OB star 
capable of photoionizing the nebula. 
An additional piece of circumstantial evidence against IC 2144 being
an emission nebula is that it was not detected in a VLA
survey of candidate \HII\ regions at 5 GHz by Fich (1993), implying an upper limit to
free-free emission from ionized gas of $< 3$ mJy, versus 50-$10^4$ mJy
for typical \HII\ regions detected in that survey.
Together these facts provide strong evidence that the optical nebula is seen in
reflected starlight.

The centrosymmetric polarization pattern we
observe now establishes unambiguously that IC 2144 is primarily a
reflection nebula illuminated by MWC 778.  Our measurement directly
confirms this at near-infrared wavelengths, and in combination with the
spectral information from HV08 (particularly the lack of [\OIII]
emission and the width of the H$\alpha$ line) we conclude that the optical 
nebula almost certainly has its origins in
reflection as well. It would be straightforward to confirm this through 
optical imaging polarimetry of IC 2144, or flux-calibrated narrow-band
imaging in H$\alpha$ and an adjacent continuum filter.

The mechanism creating the extended mid-infrared light is less clear. Scattered light may still be 
a significant, perhaps dominant, effect, but we cannot rule out contributions by thermal emission from transiently heated dust grains or fluorescent
emission from polycyclic aromatic hydrocarbons.  Mid-IR
spectroscopy of the extended nebula would allow the nature of the
emission there to be determined.

\subsection{What is the nature of the central source?}

Based on its spectrum showing many emission lines indicative of accretion plus absorption from lithium (HV08), the presence of extensive surrounding nebulosity, and the
fact that its SED is still rising past 60 microns, MWC 778 appears to be a young star. 
We can attempt to determine 
its nature more precisely by inferring its position along pre-main-sequence evolutionary tracks based on its observed properties, but
the large uncertainty in its distance prevents us from unambiguously placing MWC 778 on such tracks. 

Using their estimated 1 kpc distance, HV08 integrated MWC 778's SED to obtain a bolometric luminosity of $\sim510 L_\odot$, in agreement with previous 
estimates (see \S 2). However, the presence of the disk
implies that the source suffers more extinction than was assumed by HV08.
In fact the total extinction around
the star must depend on the line of sight, from
hundreds of magnitudes directly through the disk midplane to only a few magnitudes or less for
light scattering out along the polar cavities. 
(Thus our statement in  \S \ref{results_disk} of $A_V > 50$ in the midplane is not
incompatible with HV08's estimated $A_V \sim 3$ in total integrated light, since the latter was derived from 
the relative strengths of emission lines seen in light that presumably traveled through low extinction regions.)
\citet{Whitney:2003p2594} computed model SEDs for
class I YSOs surrounded by disks plus bipolar envelopes (like MWC 778), and found that
over a wide range of inclinations the true bolometric luminosity was
typically $1.5-2\times$ the observed luminosity. If we correct for MWC
778's non-isotropic radiation using this factor, its bolometric
luminosity increases to $750 L_\odot$ or more.  
Because the total amount of extinction is uncertain, this
should be taken as a lower limit: Luminosity correction factors of
$10\times$ or higher have been derived for other disk+bipolar envelope
systems (e.g. Perrin et al. 2007), so it seems possible that MWC 778's
luminosity could be as high as several thousand $L_\odot$. 

When combined with the F-G
spectral type claimed by HV08, 
a luminosity of $\geq 750 ~L_\odot$  
presents a conundrum, as it is
far larger than that of any other known IMTTS. That luminosity, along with an effective temperature of $\sim 7000$ K (for an F star), 
would place MWC 778 well above the birthline in the HR diagram expected for any intermediate mass stars
\citep{Palla:1990p180,Palla:2005p158}.
Although the exact position of
this stellar birthline varies depending on the protostellar mass accretion 
rate, there is no evidence that such rates are ever high enough to
allow for a 750 $L_\odot$ F star; the birthline appears to be a firm
boundary on allowable stellar parameters (Palla 2005).
We thus must reject the notion that MWC 778 could really be a single F or G star with $L\geq 750~ L_\odot$.

A similar conclusion can be reached if we use the calibrations between Br $\gamma$ and Pa $\beta$
luminosities and accretion luminosity given by Calvet et~al.\ (2000, 2004). The observed line
luminosities of MWC 778 from HV08 yield a lower limit to the accretion luminosity of $\sim 12
L_\odot$. This is several times larger than accretion luminosities measured for even the most
luminous objects in the Calvet et al. (2004) sample of IMTTS, again indicating an alternate
explanation for MWC 778 must be sought.   

The high total and accretion luminosities could instead be consistent with MWC 778 being a Herbig Be
star around $\sim 5~M_\odot$.  Indeed, most claimed spectral types for MWC 778 prior to HV08
were of type B (B1?e, Vieira et~al.\ 2003; BQ[~], Su\'{a}rez et~al.\ 2006), and similar
emission-line-dominated spectra are seen for Be stars such as LkHa 101 or MWC 1080.  
In HV08's data, MWC 778 does not show any absorption lines such as \HeI\ which would indicate the 
presence of an OB star (though emission features could potentially hide such lines). 
The fact that IC 2144 is a reflection nebula, not a photoionized HII region, places an upper limit on the amount of
ionizing flux which could be present, inconsistent with spectral types hotter than B1, but we
cannot rule out a mid-B star on this basis.  Nevertheless,
HV08 unambiguously detected optical absorption lines such as Ca I 6102.72 \AA, Fe I 6136.63/6137.70
and Li 6707 \AA, which require cooler temperatures consistent with F-G spectral type and
inconsistent with classification as a Be. The $v \sin i$ of those features is $\sim 30$ km s$^{-1}$,
a typical value for an F5 star but much less than would be expected for any B or A star
seen at high inclination ($v \sin i > 100-200$km s$^{-1}$). The observed features are shallow,
suggesting veiling, but their detection is clear and any model of MWC 778 must account for 
them.

There are (at least) three possible explanations for the apparent discrepancy between MWC 778's
large bolometric luminosity and the presence of metallic absorption lines that suggest a late
spectral type:  (1) The first and simplest explanation is that the distance is not 1 kpc, hence the
luminosity is wrong; (2) alternately, the absorption lines reported by HV08 may arise from the
atmosphere of the accretion disk, rather than the stellar photosphere, hence the spectral type is
wrong; (3) the source may be a binary system consisting of a B-type star and an F-type star.  We now
consider each of these in turn. 

If we invoke a smaller distance, how close could it plausibly be?  A potential lower limit comes from the fact
that based on projected stellar densities, IC 2144 appears to be more distant than the cloud containing RR Tau at 380 pc
(see HV08 \S 1).  If the distance to MWC 778 is actually 500 pc, rather than 1 kpc, the luminosity decreases to ~180
$L_\odot$ (and the disk radius would decrease to about 600 AU).  The observed absorption line depths are shallow,
consistent with additional continuum veiling $\sim 3\times$ greater than the intrinsic stellar photosphere (see Figure 7 in HV08). This would reduce the actual
stellar luminosity to about 45 $L_\odot$, approximately in line with that expected for a
F-type IMTTS or a Herbig Ae star near the birth line.  

But a non-stellar origin for the absorption spectra may offer an 
equally plausible means of reconciling the observed luminosity and spectrum. 
If MWC 778 is an early to mid-type Be star, surrounded by a protoplanetary disk with a nearly edge-on inclination, light reaching us must propagate
nearly parallel to the disk's surface, and indeed must scatter from dust in the optically thin surface layers to produce
the observed polarization.  Since this propagation path passes through gas as well as dust, the former will produce absorption lines in the spectra.  
To impose F/G type absorption lines on an originally near-featureless Be star
spectrum requires gas temperatures of several thousand degrees, several times larger than the expected dust destruction temperature.
However, \citet{Kamp:2004p2484} showed that the gas and dust temperatures decouple in a protoplanetary disk's upper
atmosphere, and that the gas between the dust disk's surface and a few AU above the disk can be as hot as $10^4$ K at distances of tens of AU
from a T Tauri star.  A similar disk atmosphere model would need to be calculated for a Be star to investigate whether a sufficient
optical depth of gas at $6000 - 7000 $K could be available in the disk atmosphere of MWC 778 to produce the observed lines. 
In this scenario, the 30 km$^{-1}$ widths of the absorption lines would be due to the superposition
of light scattered
toward us from multiple portions of the rotating disk, each with its own Keplerian velocity relative
to us.

Finally, it is possible that MWC 778 is a binary system, consisting of
a B star and an F star, similar to Z CMa (e.g., Whitney et al.\ 1993).
The high resolution optical spectra obtained by HV08 reveal
no evidence of velocity shifts in the absorption
lines due to orbital motions, but the signal-to-noise ratio of these features is low, and the
upper limit on any velocity shifts is large ($< 10$ km/s; Herbig,
private communication).  To investigate how detectable binarity would
be given HV08's two spectroscopic
observations separated by about one year, we performed Monte Carlo simulations of binary systems with a $4
M_\sun$ primary and an inclination of $70 \deg$, for companion masses
ranging from the substellar mass limit to $4 M_\sun$, under the
assumption that the observed absorption lines arise in the atmosphere
of the lower-mass component. These simulations indicate that 
the observations of HV08 
would have detected
line shifts of $\geq$ 10 km s$^{-1}$ with a probability of $\sim0.9$ if there existed any companion with a semi-major
axis $a < 5$ AU, roughly independent
of companion mass. Therefore we can
rule out such a close binary in MWC 778. However, the existing
observations have only a probability of $\sim 0.5$ of detecting any binary
companion with $5 < a < 10$ AU, and are completely insensitive to
binaries with larger semi-major axes $a > 10$ AU.

Examination of the evolutionary tracks by Palla \& Stahler (1990)
indicate that a binary system consisting of a late B star and an early
F star in which both stars sit on the birth line could just produce
the estimated luminosity for MWC 778. Such a system would have
approximately the correct secondary-to-primary flux ratio ($\sim0.2$, based
on HV08's observation that the F-type metal absorption lines are
about a factor of 3-4 weaker than expected from a normal F star).
However this is the largest luminosity such a binary could have; as
soon as the stars age and move off the birth line, the total
luminosity drops precipitously. For an age of 1 Myr, such a system
would be expected to have a total luminosity of only $\sim 170 L_\sun$.
In short, a B star + F star binary system could explain the observed
properties of MWC 778 only if either the system is extremely young ($<
1$ Myr) or, as above, it is much closer than 1 kpc in addition to
being a binary.

\subsection{What is the relationship between the inner disk and the outer spiral?}

HV08 conjectured that the ``reverse question mark''-shaped dark lane they observed in H$\alpha$ might be due to
dust obscuration in front of the background nebula.  We now see that the optical dark lane corresponds 
to the un-illuminated region adjacent to the spiral nebula seen 
in our infrared data. 
Given the much greater penetrating power of 10-20 $\mu$m radiation
compared to visible light, producing the observed spiral pattern with
 a foreground dust cloud would require that cloud to both be very dense (to provide sufficient opacity at $Q$ band) and
have very sharply defined edges (so that the apparent spiral does not change in position or size with wavelength). 
That scenario is not impossible, but would require some fine-tuning of the dust cloud's properties. 
We may instead be looking at an actual physical structure rather than a line-of-sight superposition.  
In  that case, it remains unclear what the spiral's origin might be. 

Spiral density waves within circumstellar disks have now
been observed around many YSOs (e.g. AB Aur, \citealp{Grady:1999p121}; HD 141569, \citealp{Clampin:2003p118}; HD 142527, \citealp{Fukagawa:2003p241}), and 
are widely interpreted to be caused by the gravitational effects of orbiting companions.  Since MWC 778 may well be a binary system, this seems at first to be an attractive explanation. 
But MWC 778's spiral differs from those seen around typical Herbig Ae/Be stars in several significant ways. First, it is much
larger. Spiral structures, such as in the disks around AB Aur and HD 141569, are typically observed on a scale of
100-300 AU. In contrast, if MWC 778 truly is about 1 kpc distant, then the observed spiral is at least
12,000 AU across!  
Even more significantly, the spiral is not part of the circumstellar
disk plane but instead appears to be
oriented perpendicular to it. 
It is highly unlikely that a
binary system whose orbital plane is aligned with the observed disk
could produce a spiral with this orientation. 

Instead of being part of the disk, the inner portions of the spiral apparently connect to MWC 778's polar
axis, aligned with the bipolar cavities there.
For some YSOs, precessing outflows have carved corkscrew-like bipolar
cavities in their envelopes.
Yet invoking outflows to carve the spiral around MWC 778 
would require sufficient precession to turn the 
entire inner disk almost end-over-end in an impossibly short period of time ($\sim 500$ years, for an observed size of 12000 AU and typical outflow velocities of 100 km/s). It is hard
to see how this could be the case. 

An alternate explanation for the spiral would be to invoke turbulence in molecular clouds, which is believed to create
localized overdensities that can become unstable and collapse, leading to star formation. The curving nebula might then represent
a residual filament of compressed material, which gave birth to the star but has not yet dispersed. A similar explanation has 
been suggested for several binary YSOs which appear to be linked by bridges of material: e.g.,  LkH$\alpha$ 225
\citep{Perrin:2004p1835} or SR 24 \citep{2002AAS...201.6607P}.  Such a scenario would imply that 
the roughly symmetrical spiral appearance around MWC 778 is just a matter of chance.

\section{Conclusions}
\label{concl_sect}

Near-infrared polarimetric and mid-infrared imaging observations of
the young stellar object MWC 778 have revealed a
nearly-edge-on circumstellar disk plus bipolar outflow
cavities, confirming the hypothesis advanced by HV08 on the basis
of their spectra of the star and its nearby nebulosity.  
The images also reveal a large and intriguing spiral structure
surrounding the star, prominent in all of the H$\alpha$, near- and
mid-infrared images.  Because it appears to be oriented perpendicularly to the
circumstellar disk plane, it is difficult to explain this as a spiral density wave similar to those seen in disks around
many other young stars.  The nature of this feature is currently unknown.

The presence of a disk suggests that previous values for the
luminosity of the source may be underestimates; after correction for light blocked by the
disk, the total bolometric
luminosity at the assumed distance of 1 kpc now becomes at least $\sim 750 L_\odot$. 
This leads to a paradox:
that luminosity is far too
large for the late-type spectral type derived
by HV08 based on absorption features seen in the optical.
Even the most luminous young F- and G-type stars, the
intermediate mass T Tauri stars, are not expected to have
luminosities in excess of $\sim 100 L_\odot$. 

We have suggested several possibilities to
resolve this discrepancy: (1) the distance to MWC 778 may be vastly
over-estimated, and it is really an IMTTS at a distance of a few hundred pc instead of 1 kpc; 
(2) the F-G-type absorption features may arise from
the atmosphere of the circumstellar disk rather than the stellar
photosphere, in which case the star itself becomes most plausibly a young Be star; 
(3) the system may actually be a binary composed of a B star and an F star. 
The currently available data do not allow us to determine
which of these possibilities is correct, and in fact they are not mutually exclusive.

How might we test any of these hypotheses for the central source? The highest priority may be further spectroscopic
observations, both to obtain increased temporal coverage in search of
radial velocity variations from orbital motion, and to obtain
higher signal-to-noise on the various absorption features to better
constrain the apparent spectral type. Given the faintness of this
source in the optical ($V\sim 14$) such observations will be
challenging.  Further observations of the
disk may provide indirect constraints on the stellar properties. Mapping the region in molecular
emission lines (e.g. CO) could detect Keplerian rotation in the disk, providing kinematic constraints on the
mass of the central object. Tighter constraints on the maximum amount
of ionizing radiation present could potentially be obtained through
flux-calibrated H$\alpha$ imaging---but such measurements would need to account for accretion
luminosity and would be fraught with difficulties.  Lastly, deeper wide-field imaging in the infrared should
be obtained in order to search for additional nearby embedded sources
that might be present; if any associated young but less deeply-embedded sources
can be found, their properties may provide complimentary constraints on the
distance.

High contrast imaging observations are a powerful tool for investigating circumstellar
environments and the physical processes at work therein, 
particularly when multiwavelength observations are
available for analysis (e.g. \citealp{2007prpl.conf..523W,Pinte:2008p2580}). The spatially resolved observations
presented here are a first step towards applying such a ``pan-chromatic''
approach to the study of MWC 778's disk.  
But detailed
investigations of disk structure are unlikely to be fruitful without a better
understanding of the illuminating source and the distance to the
system.
As is so often said, establishing accurate distances remains one of the most
challenging tasks in astronomy---yet doing just that is likely 
a necessary first step in 
conclusively establishing the nature of MWC 778.

\acknowledgements

WDV would like to thank George Herbig for calling his attention to
this source and for numerous discussions regarding its nature, and also  
thanks Goeran Sandell for his insight regarding Herbig Ae/Be stars.
MDP thanks John Monnier for useful discussions. These
observations were made possible by the dedicated and hard working staffs of the Lick
and Gemini Observatories, and in particular MDP is most grateful for expert assistance from Elinor Gates, R.
Scott Fisher, and Kevin Volk.

The data presented here were obtained as part of a thesis project
by MDP, who was supported at that time by a NASA Michelson Graduate
Fellowship administered by JPL, and is is now supported by a NSF
Astronomy \& Astrophysics Postdoctoral Fellowship. JRG \& MDP
were supported in part by the National Science Foundation Science and
Technology Center for Adaptive Optics, managed by the University of
California at Santa Cruz under cooperative agreement No. AST-9876783.
This paper is based in part on
observations obtained at the Gemini Observatory, which is operated by the
Association of Universities for Research in Astronomy, Inc., under a cooperative agreement
with the NSF on behalf of the Gemini partnership: the National Science Foundation (United
States), the Science and Technology Facilities Council (United Kingdom), the
National Research Council (Canada), CONICYT (Chile), the Australian Research Council
(Australia), Minist\'erio da Ci\^encia e Tecnologia (Brazil) and SECYT (Argentina). The Gemini observing time was made
available through a time trade arrangement between Gemini and the W. M. Keck Observatory, which is operated as a
scientific partnership among the California Institute of Technology, the University of California and the National
Aeronautics and Space Administration, and was made possible by the generous financial support of the W.M.
Keck Foundation.
The authors wish to recognize and acknowledge the very significant cultural role and reverence that the summit of Mauna
Kea has always had within the indigenous Hawaiian community.  We are most fortunate to have the opportunity to conduct
observations from this mountain.

\end{document}